\documentclass[entropy,article,accept,oneauthor,pdftex,12pt,a4paper]{mdpi}

\lastpage{x}
\doinum{10.3390/------}
\pubvolume{xx}
\pubyear{2013}
\history{Received: xx / Accepted: xx / Published: xx}
\pdfoutput=1

\usepackage{amsthm, amssymb,enumerate,framed,listings,multirow,pdftricks,subcaption,times,xcolor}

\colorlet{shadecolor}{gray!15}

\hypersetup{
    bookmarks=true,         
    unicode=false,          
    pdftoolbar=true,        
    pdfmenubar=true,        
    pdffitwindow=false,     
    pdfstartview={FitH},    
    pdftitle={My title},    
    pdfauthor={Author},     
    pdfsubject={Subject},   
    pdfcreator={Creator},   
    pdfproducer={Producer}, 
    pdfkeywords={keyword1} {key2} {key3}, 
    pdfnewwindow=true,      
    colorlinks=false,       
    linkcolor=black,          
    citecolor=black,        
    filecolor=black,      
    urlcolor=black           
}

\def\Y_#1{\vect{y}_{\!#1}}

\def\E{\mathbb{E}}

\def\divergence{\operatorname{div}}

\newcommand{\vect}[1]{\boldsymbol{#1}}

\newcommand{\eqdef}{\stackrel{\mbox{\tiny def}}{=}}

\newtheorem{theorem}{Theorem}[section]
\newtheorem{thm}{Theorem}[section]

\theoremstyle{definition}

\newtheorem{algorithm}[theorem]{Algorithm}

\theoremstyle{remark}

\newenvironment{algo}
  {\begin{shaded}\begin{algorithm}}
  {\end{algorithm}\end{shaded}}

\def\E{\mathbb{E}}

\Title{On Metropolis Integrators for Molecular Dynamics}

\Author{Nawaf Bou-Rabee $^{1,}$*}

\address[1]{$^{1}$ Department of Mathematical Sciences, Rutgers University--Camden, 311 N 5th Street, Camden, NJ 08102, USA}

\corres{nawaf.bourabee@rutgers.edu}

\abstract{
This paper invites the reader to experiment with an easy-to-use MATLAB \cite{MATLAB:2012} implementation of Metropolis integrators for Molecular Dynamics (MD) simulation  \cite{BoVa2010,BoVa2012}.  These integrators are analysis-based, in the sense that they can rigorously simulate dynamics along an infinitely long MD trajectory.  Among explicit integrators  for MD, they seem to be the only ones that satisfy the fundamental requirement of stability. The schemes can handle stiff or hard-core potentials, and are straightforward to set up, apply and extend to new situations.  Potential pitfalls in high dimension are discussed, and tricks for mitigation are given.  
}



\begin{document}

\lstset{language=Matlab}          
\lstset{numbers=left, numberstyle=\tiny, stepnumber=1, numbersep=5pt, keywordstyle=\bfseries,basicstyle=\ttfamily,  frame=lines}

\section{Introduction}

\begin{figure}[ht!]
\includegraphics[width=1\textwidth]{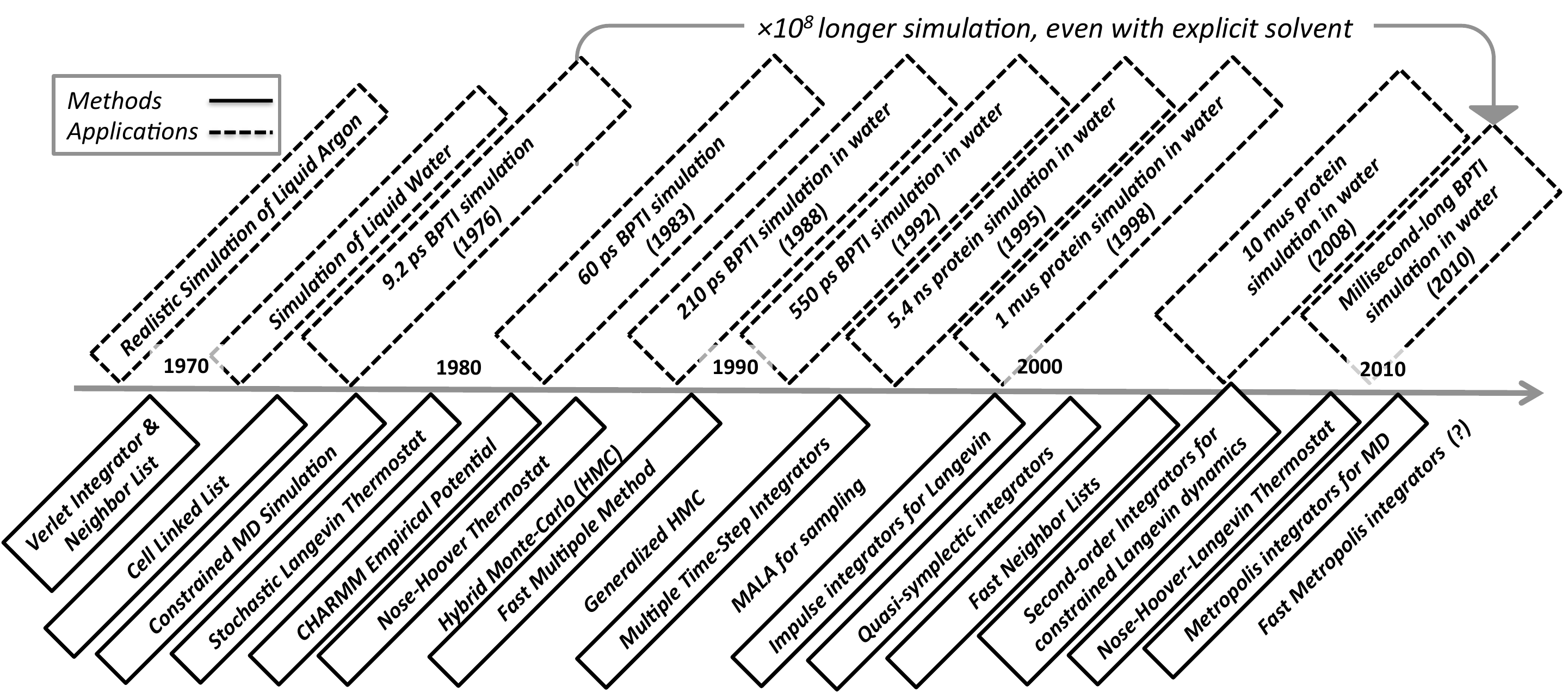} 
\caption{ \small {\bf  Timeline of Selected MD Developments.}  A timeline of the evolution of MD simulation methodology and applications.  The applications highlighted are simulations of liquid argon \cite{rahman1964correlations}, water \cite{rahman1971molecular},  protein dynamics without solvent \cite{mccammon1977dynamics,van1977algorithms}, and biopolymer dynamics with solvent \cite{levitt1983molecular,levitt1988accurate,daggett1992model,li1995investigation,duan1998pathways,freddolino2008ten,shaw2010atomic}.  The methods include the following upgrades to MD simulation: Verlet integrator and neighbor lists \cite{Ve1967}, cell linked list \cite{quentrec1973new}, SHAKE integrator for constraints \cite{RyCiBe1977}, stochastic heat baths via Langevin dynamics \cite{ScSt1978, BrBrKa1984}, library of empirical potentials \cite{brooks1983charmm}, extended/deterministic heat bath via Nos\'{e}-Hoover dynamics \cite{No1984,Ho1985}, fast multipole method \cite{greengard1987fast}, multiple time-steps \cite{TuBeMa1992}, Langevin splitting methods \cite{SkIz2002, MiTr2003}, (fast) combined neighbor and cell lists \cite{yao2004improved}, second-order Langevin integrators for constrained systems \cite{VaCi2006}, and stochastic Nos\'{e}-Hoover Langevin thermostat \cite{SaChDe2007,LeNoTh2009,LeRe2009}.
 }
\label{fig:timeline}
\end{figure}

Molecular Dynamics (MD) simulation refers to the time integration of Hamilton's equations often coupled to a heat or pressure bath  \cite{AlTi1987, FrSm2002, rapaport2004art,tuckerman2008statistical, schlick2010molecular}. From its early use in computing 
 equilibrium dynamics of homogeneous molecular systems \cite{rahman1964correlations,Ve1967,alder1967velocity,alder1970studies,harp1970time,rahman1971molecular,stillinger1974improved,stillinger1980water}
and pico to nanoscale protein dynamics \cite{mccammon1977dynamics,van1977algorithms,mccammon1980simulation,van1982protein,karplus1983dynamics,van1990computer,karplus2002molecular,case2002molecular,adcock2006molecular,van2008molecular}, the method has evolved into a general purpose tool for simulating statistical properties of heterogeneous molecular systems \cite{kapral2005molecular}.  Accessible time horizons have increased remarkably:  the timeline in Figure~\ref{fig:timeline} attempts to capture this nearly billion-fold improvement in capability over the last forty or so years. Near future applications include micro to milliscale simulations of biomolecular processes like protein folding, ligand binding, membrane transport, and  biopolymer conformational changes \cite{scheraga2007protein,dror2012biomolecular,lane2012milliseconds}.    In addition, atomistic MD simulations are used more sparingly in multiscale models \cite{nielsen2004coarse, tozzini2005coarse, clementi2008coarse, sherwood2008multiscale, weinan2011principles} and rare event simulation such as the finite temperature string method and milestoning \cite{EVa2004,VaVe2009A, VaVe2009B,EVa2010}.   Given this continuous development and generalization of MD, it is not a stretch to suppose that MD will play a transformative role in medicine, technology, and education in the twenty-first century.

In its standard form, the method inputs a random initial condition, fudge factors, physical and numerical parameters; and, outputs a long discrete path of the molecular system.  Statistical quantities, like velocity correlation or mean radius of gyration, are usually computed online, i.e., as points along this trajectory are produced.   MD simulation is built atop a forward Euler-like integrator that requires a single interactomic force field evaluation per step.  Even though MD sounds quite simple, software implementations of MD are typically optimized for performance \cite{brooks1983charmm,nelson1996namd,scott1999gromos}, and as a side effect, obscure this simplicity and make it cumbersome for newcomers to learn, modify, test and propose enhancements.

Besides this steep learning curve, due to the interplay between stochastic Brownian and interatomic forces, current MD integrators are unable to stably produce long trajectories.  This is a well known difficulty with explicit integrators for nonlinear diffusions \cite{Ta2002, HiMaSt2002, MiTr2005,Hi2011,HuJeKl2012}.   Recently, a probabilistic solution to this problem was proposed that challenges the traditional notion that Monte-Carlo methods and MD have disjoint aims: the former strictly samples probability distributions and the latter estimates dynamics. The basic idea is to combine a standard MD integrator with a Metropolis-Hastings algorithm targeted to the Gibbs-Boltzmann distribution \cite{AkBoRe2009,BoVa2010, BoVa2012}.  Because the scheme is a Monte-Carlo method it exactly preserves the Gibbs-Boltzmann distribution \cite{AkBoRe2009,BoVa2010}.  This important property implies numerical stability over long-time simulations.  In addition, a Metropolized integrator is also accurate on finite time intervals \cite{BoVa2010}, and so, even though a Metropolized integrator involves a Monte-Carlo step, its aim and philosophy are very different from Monte-Carlo methods whose only goal is to sample a target distribution with no concern for the dynamics \cite{MeRoRoTeTe1953,Ha1970, RoDoFr1978, DuKePeRo1987, Ho1991, KePe2001, Li2008, AkRe2008, AkBoRe2009, LeRoSt2010A}.

Motivated by these issues, this paper builds a software implementation of a `Metropolis integrator' and applies it to a homogeneous molecular system.  The algorithms are introduced in a step-by-step fashion.  The software version of the algorithm is written in the latest version of MATLAB with plenty of comments, variables that are descriptively named, and operations that can be easily translated into mathematical expressions \cite{MATLAB:2012}.   Since MATLAB is widely available, this design ensures the software will be easy-to-use and cross-platform.  The following MATLAB-specific file formats will be used.
\begin{description}
\item[(F1) MATLAB Script \& Function] files are written in the MATLAB language, and can be run from the MATLAB command line without ever compiling them.   
\item[(F2) MATLAB Executable (MEX)] files are written in the `C' language and compiled using the MATLAB {\tt mex} function.  The resulting executable is comparable in efficiency to a `C' code and can be called directly from the MATLAB command line. We will use MEX-files for performance-critical routines \cite{mexfiles2013}.  
\item[(F3) MATLAB Binary (MAT)] files will be used to store simulation data.
\end{description}

The paper is organized as follows.  We begin with some mathematical background in \S\ref{preliminaries}, followed by an algorithmic introduction to time integrators in MD simulation and a statistical analysis of a long trajectory of a Lennard-Jones fluid in \S\ref{algorithm}.  The paper discusses some tricks to get the integrator to scale well in high dimension in \S\ref{scaling}.   We close the paper with some steps for future development of Metropolis integrators in \S\ref{conclusion}.


\section{Mathematical Background}   \label{preliminaries}

\subsection{Bath--Free Dynamics}

MD is based on Hamilton's equations for a Hamiltonian $H : \mathbb{R}^{2 d} \to \mathbb{R}$: 
\begin{equation} \label{hamiltons}
 \dot{\vect{z}}(t) = \vect{J} \nabla H(\vect{z}(t)) \;,~~ \vect{z}(0) \in \mathbb{R}^{2 d} \;,
\end{equation}
where $\vect{z}(t) = (\vect{q}(t), \vect{p}(t))$ is a vector of molecular positions $\vect{q}(t) \in \mathbb{R}^d$ and momenta $\vect{p}(t) \in \mathbb{R}^d$, and
$\vect{J}$ is the $2 d \times 2d$ skew-symmetric matrix defined as: \[ 
\vect{J} = \begin{pmatrix} \vect{0}_{d \times d} & \vect{I}_{d \times d} \\
- \vect{I}_{d \times d} & \vect{0}_{d \times d} \end{pmatrix} \;.
\] The Hamiltonian $H( \vect{z} )$ represents the total energy of the molecular system, and is typically  `separable' meaning that it can be written as: \[
H(\vect{z} ) = K(\vect{p}) + U(\vect{q}) \;, ~~ \vect{z} = (\vect{q}, \vect{p}) \;,
\]  where $K(\vect{p})$ and $U(\vect{q})$ are the kinetic and potential energy functions, respectively \cite{MaRa1999}.   In MD, the kinetic energy function is a positive definite quadratic form, and the potential energy function involves `fudge factors' determined from experimental or quantum mechanical studies of pieces of the molecular system of interest \cite{brooks1983charmm}.  The accuracy of the resulting energy function must be systematically verified by comparing MD simulation data to experimental data \cite{van1998validation}.  The flow that \eqref{hamiltons} determines has the following structure:
\begin{description}
\item[(S1)] volume-preserving (since the vector-field in \eqref{hamiltons} is divergenceless); and,
\item[(S2)] energy-preserving (since $\vect{J}$ is skew-symmetric \& constant).
\end{description}
Explicit {\em symplectic integrators} -- like the Verlet scheme -- exploit these properties to obtain long-time stable schemes for Hamilton's equations \cite{LeRe2004,HaLuWa2010}.

\subsection{Governing Stochastic Dynamics}

In order to mimic experimental conditions,  \eqref{hamiltons} is often coupled to a bath that puts the system at constant temperature and/or pressure.   The
standard way to do this is to assume that the system with bath is governed by a stochastic ordinary differential equation (SDE) of the type:  \begin{equation} \label{sde} \boxed{
d \vect{Y}(t) = \underset{\text{deterministic drift}}{\underbrace{\vect{A}(\vect{Y}(t)) dt}} +  \underset{\text{heat bath}}{\underbrace{(\divergence \vect{D})(\vect{Y}(t))  dt + \sqrt{2 kT} \vect{B}(\vect{Y}(t)) d \vect{W}(t)}}
}
\end{equation}
Here, we have introduced the following notation.
\begin{center}
\begin{tabular}{cc}
\hline
$\vect{Y}(t) \in \mathbb{R}^{n}$ &  state of the (extended) system \\
$\vect{A}(\vect{x}) \in \mathbb{R}^n$ & deterministic drift vector field \\
$\vect{B}(\vect{x}) \in \mathbb{R}^{n \times n}$ & noise-coefficient matrix \\
$\vect{D}(\vect{x}) \in \mathbb{R}^{n \times n}$ & diffusion matrix \\
$\vect{W}(t) \in \mathbb{R}^n$ &  $n$-dimensional Brownian motion  \\
$kT$ &  temperature factor \\
\hline
\end{tabular}
\end{center} 
The $n \times n$ {\em diffusion matrix} $\vect{D}(\vect{x})$ is defined in terms of the noise coefficient matrix $\vect{B}(\vect{x})$ as: \begin{equation} \label{diffusion}
\vect{D}( \vect{x} ) \eqdef kT \vect{B}( \vect{x} ) \vect{B}( \vect{x} )^T \;, ~~ \text{for all $\vect{x} \in \mathbb{R}^n$} \;,
\end{equation}
where $\vect{B}( \vect{x} )^T$ denotes the transpose of the real matrix $\vect{B}( \vect{x} )$.    The diffusion matrix is symmetric and nonnegative definite.
Depending on the particular bath that is used, the dimension $n$ of $\vect{Y}(t)$ in \eqref{sde} is related to the dimension $2 d$ of $\vect{z}(t)$ in \eqref{hamiltons} by the inequality: $n \ge 2 d$.  For example, in Nos\'{e}-Hoover Langevin dynamics a single bath degree of freedom is added to \eqref{hamiltons} so that $n=2 d +1$, while in Langevin dynamics the effect of the bath is modeled by added friction and Brownian forces that keep $n = 2d$.    Throughout this work, MD simulation refers to time integration of \eqref{sde} from a random initial condition.

The equations \eqref{sde} generate a stochastic process $\vect{Y}(t)$ that is a {\em Markov diffusion} process.   We assume that this diffusion process admits a {\em stationary} distribution $\mu(d \vect{x})$, i.e., a probability distribution preserved by the dynamics \cite{IkWa1989,klebaner2005introduction}.   We denote by $\nu(\vect{x})$ the density of this distribution. Even though the diffusion matrix \eqref{diffusion} is not necessarily positive definite, one can use  {\em H\"{o}rmander's condition} to prove that the process $\vect{Y}(t)$ is an ergodic process with a unique stationary distribution \cite{MenTw1996,DaZa1996}.  By the ergodic theorem, it then follows that:
\begin{equation}
  \label{eq:ergo1}
 \frac{1}{T} \int_0^T f( \vect{Y}(t) ) dt \to \int_{\mathbb{R}^n} f( \vect{x} ) \nu(\vect{x}) d \vect{x} \;, ~~\text{as}~~ T \to \infty,
\end{equation}
where $f(\vect{x})$ is a suitable test function.

The evolution of the probability density of the law of $\vect{Y}(t)$ at time $t$, $\rho(t, \vect{x})$, satisfies the Fokker-Planck equation: \begin{equation} \label{fp}
- \frac{\partial \rho}{\partial t} + L \rho = 0 \;, 
\end{equation} where $\rho(0, \cdot)$ is the density of the initial distribution $\vect{Y}(0) \sim \rho(0, \cdot)$, and $\mathcal{L}$ is defined as the following 
second-order partial differential operator: \[
( L f)(\vect{x}) \eqdef \divergence\left(  \divergence(\vect{D}(\vect{x}) f(\vect{x})) - \vect{A}(\vect{x}) f(\vect{x}) \right) \;.
\]  Since $\mu(d \vect{x} ) = \nu(\vect{x}) d \vect{x}$ is a  stationary distribution of $\vect{Y}(t)$, the probability density $\nu(\vect{x})$ is a steady-state solution of \eqref{fp}: \begin{equation} \label{stationarity}
( L \nu)( \vect{x}) = 0 \;.
\end{equation}
Define the {\em probability current} as the vector field: \begin{equation} \label{current}
\vect{j}(\vect{x}) \eqdef \divergence(  \vect{D}( \vect{x}) \nu( \vect{x}) ) - \vect{A}(\vect{x}) \nu(\vect{x}) 
\end{equation}
The stationarity condition \eqref{stationarity} implies that $\vect{j}(\vect{x})$ is divergenceless.   In the zero-current case, the diffusion process $\vect{Y}(t)$ is {\em reversible} and the stationary density $\nu(\vect{x})$ is called the equilibrium probability density of the diffusion \cite{HaPa1986}.

In this case, the operator $L$ is self-adjoint, in the sense that:
\begin{equation}
\label{eq:idb}
\langle Lf, g \rangle_{\nu} = \langle f, L g \rangle_{\nu} \qquad \text {for all suitable test functions $f, g$}  \;,
\end{equation}
where $\langle \cdot , \cdot  \rangle_{\nu}$ denotes an $L^2$ inner product weighted by the density $\nu(x)$.   This property implies that the diffusion is $\nu$-symmetric~\cite{Ke1978}:
\begin{equation}
  \label{eq:db}
  \nu(\vect{x}) p_t(\vect{x},\vect{y}) = \nu(\vect{y}) p_t(\vect{y},\vect{x}) \qquad \text {for all $t>0$} \;,
\end{equation}
where $p_t(\vect{x},\vect{y})$ denotes the transition probability density of $\vect{Y}(t)$.  Indeed~\eqref{eq:idb} is simply an infinitesimal version of  \eqref{eq:db}, which is referred to as the detailed balance condition.   In the self-adjoint case, the drift is uniquely determined by the diffusion matrix and the stationary density $\nu(\vect{x})$: \[
\vect{j}(\vect{x}) = \vect{0} \implies  \vect{A}(\vect{x}) = \frac{1}{\nu(\vect{x})}  \divergence(  \vect{D}( \vect{x}) \nu( \vect{x}) ) \;.
\]
Long-time stable explicit schemes adapted to this structure have been recently developed \cite{BoDoVa2013}.

\subsection{Splitting Approach to MD Simulation}

We are now in position to explain our approach to deriving a long-time stable scheme for \eqref{sde}.  Crucial to our approach is that in  MD simulation we usually have a formula for a function proportional to the stationary density $\nu(\vect{x})$. Following \cite{BoOw2010}, we can split \eqref{sde} into:  \begin{equation} \label{sdesplit1} 
d \vect{Y} = - \vect{D}(\vect{Y}) \nabla H_{\nu}(\vect{Y}) dt +  \divergence \vect{D}(\vect{Y}) dt + \sqrt{2 kT} B(\vect{Y}) dW 
\end{equation} \begin{equation} \label{sdesplit2} 
\dot{\vect{Y}} = \vect{A}(\vect{Y}) + \vect{D}(\vect{Y}) \nabla H_{\nu}(\vect{Y})  
\end{equation} where we have introduced $H_{\nu}(\vect{x}) = - ( \log \nu )(\vect{x})$.   An {\em exact splitting method} preserves $\mu(d \vect{x})$. It is formed by taking the exact solution (in law) of \eqref{sdesplit1} in composition with the exact flow of \eqref{sdesplit2}.   The process produced by \eqref{sdesplit1} is self-adjoint with respect to $\nu(\vect{x})$.  Moreover, stationarity of $\nu(\vect{x})$ implies that the flow of the ODE \eqref{sdesplit2} preserves it.   Since each step is preservative, their composition is too.

In place of the exact splitting, a Metropolis integrator can be used for \eqref{sdesplit1} \cite{BoDoVa2013}, and a measure-preserving scheme can be designed to solve the ODE  \cite{ezra2006reversible, BoVa2010}.   In \cite{BoDoVa2013}, explicit schemes are introduced for \eqref{sdesplit1} that, for the first time:  (i) sample the exact equilibrium probability density of the SDE when this density exists (i.e., whenever $\nu(\vect{x})$ is normalizable); (ii) generates a weakly accurate approximation to the solution of~\eqref{sde} at constant $kT$; (iii) acquire higher order accuracy in the small noise limit, $kT \to 0$;  and, (iv) avoid computing the divergence of the diffusion.  Compared to the methods in~\cite{BoVa2010}, the main novelty of these schemes stems from (iii) and (iv).   The resulting explicit splitting method is accurate, since it is an additive splitting of \eqref{sde}; and typically ergodic when the continuous process is ergodic \cite{BoVa2010}.

This type of splitting of \eqref{sde} is quite natural and has been used before in:  MD \cite{VaCi2006, BuPa2007}, dissipative particle dynamics  \cite{Sh2003, SeFaEsCo2006}, and simulation of inertial particles \cite{PaStZy2008}.      Other related schemes for \eqref{sde} include Br\"{u}nger-Brooks-Karplus (BBK) \cite{BrBrKa1984},  van Gunsteren and Berendsen (vGB) \cite{GuBe1982},   the Langevin-Impulse (LI) methods  \cite{SkIz2002}, and quasi-symplectic integrators \cite{MiTr2003}.   The long-time statistical properties  of this splitting was quantified in the context of globally Lipschitz potential forces in \cite{BoOw2010}.    However, for general MD force fields, none of these explicit integrators are stable.  Our framework to stabilize explicit MD integrators is the Metropolis-Hastings algorithm.

 \subsection{Metropolis-Hastings algorithm}

A Metropolis-Hastings method is a Monte-Carlo method for producing samples from a probability distribution, given a formula for a function proportional to its density \cite{MeRoRoTeTe1953,Ha1970}.    The algorithm consists of two sub-steps: firstly, a proposal move is generated according to a transition density $g( \vect{x}, \vect{y})$; and second, this proposal move is accepted or rejected with a probability:
\begin{equation}
  \alpha(\vect{x}, \vect{y}) =  1 \wedge 
  \frac{g(\vect{y},\vect{x}) \nu(\vect{y}) }
  { g(\vect{x},\vect{y}) \nu(\vect{x}) }  \;.
\end{equation}
Standard results on Metropolis-Hastings methods can be used to classify this algorithm as ergodic \cite{Nu1984, Ti1994, MenTw1996}.


\section{Algorithmic Introduction to MD Integrators}  \label{algorithm}

We now focus our discussion on Langevin dynamics of a system of $N$ atoms.  Denote by $m_j>0$ and $\vect{q}_j$ the mass and position of the $j$th atom, respectively.  The governing Langevin equation is given by: \begin{equation} \label{langevin}
\begin{cases}
 \dot{\vect{q}}_j(t) = m_j^{-1} \vect{p}_j(t) \;, \\
d\vect{p}_j(t) = - \frac{\partial U}{\partial \vect{q}_j}(\vect{q}(t)) dt - \gamma \vect{p}_j(t) dt + \sqrt{2 kT \gamma m_j}  d \vect{w}_j \;,
 \end{cases}
 ~~ j=1, \cdots, N \;,
\end{equation}
where $\vect{q} = (\vect{q}_1, \cdots, \vect{q}_{N})$ and $\vect{p} = (\vect{p}_1, \cdots, \vect{p}_{N})$ denote the positions and momenta of the particles, and $\vect{w}_j$ are independent Brownian motions.  In Langevin dynamics, positions are differentiable, and due to the irregularity of the Brownian force, momenta are just continuous but not differentiable.  The last two terms in the second equation  represent the effect of the bath.  The bath-free dynamics is a Hamiltonian system with: \[
H(\vect{q}, \vect{p}) = \sum_{j=1}^N \frac{1}{2 m_j} | \vect{p}_j |^2 + U( \vect{q} ) \;.
\]  The stationary distribution of the Langevin process has the following density: \begin{equation} \label{langevindensity}
\nu(\vect{q}, \vect{p}) = Z^{-1} \exp\left( - \frac{1}{kT} H(\vect{q}, \vect{p}) \right) \;, ~~ Z = \int \exp\left( -\frac{1}{kT}  H(\vect{q}, \vect{p}) \right) d \vect{q} d \vect{p} \;.
\end{equation}
The Langevin equation \eqref{langevin} can be put in the form of \eqref{sde} by letting $\vect{x} = (\vect{q}, \vect{p})$,  \begin{equation}
\vect{A}(\vect{x}) = \begin{pmatrix} \vect{m}^{-1} \vect{p} \\ - \nabla U(\vect{q}) - \gamma \vect{p} \end{pmatrix} \;, ~~
\vect{B} = \sqrt{\gamma} \begin{pmatrix} \vect{0} & \vect{0} \\ \vect{0}  & \vect{m}^{1/2} \end{pmatrix} \;,~~ \text{and}~~
\vect{W} = (\vect{w}_1, \cdots , \vect{w}_N) \;,
\end{equation}
where $\vect{m} = \operatorname{diag}(m_1, \cdots, m_N)$.  The splitting approach discussed in \S\ref{preliminaries} applied to Langevin dynamics \eqref{langevin} leads to the following special cases of \eqref{sdesplit1} and \eqref{sdesplit2}: \begin{equation} \label{ou}
\begin{cases}
d\vect{p}(t) =  - \gamma \vect{p}(t) dt + \sqrt{2 kT \gamma} \vect{m}^{1/2}  d \vect{W} \;,
\end{cases}
\end{equation} \begin{equation} \label{hamiltons2}
\begin{cases}
 \dot{\vect{q}}(t) = \vect{m}^{-1} \vect{p}(t) \;, \\
\dot{\vect{p}}(t) = - \nabla U(\vect{q}(t)) \;.
 \end{cases}
\end{equation}
Notice that \eqref{ou} is a linear SDE that can be exactly solved \cite[See Chapter 5]{Ev2007}.   We will use a Verlet integrator for \eqref{hamiltons2} that preserves volume and represents the energy to third-order accuracy per step.   Since the Verlet integrator does not exactly preserve energy, the composition of the two schemes does not preserve the stationary distribution with density \eqref{langevindensity}.    As a consequence of this discretization error, this scheme may either not detect properly features of the potential energy, which leads to unnoticed but large errors in dynamic quantities such as the mean first passage time, or may mishandle soft or hard-core potentials, which leads to numerical instabilities; see the numerical examples in \cite{BoDoVa2013}.  These numerical artifacts motivate adding a Metropolis accept/refusal sub-step to the integrator.    The {\em corrected} MD integrator follows.

\begin{algo}[Analysis-based MD Integrator] \label{algo:integrator}
Given the current state $(\vect{Q}_0, \vect{P}_0)$ at time $t$ the algorithm proposes a position $(\vect{Q}_1^{\star}, \vect{P}_1^{\star})$ at time $t+h$ for some time-step $h>0$ via \begin{equation}
\tag{Step 1}
\begin{cases} 
\vect{P}_{1/2} = \vect{P}_0 - \frac{h}{2} \nabla U(\vect{Q}_0) \\
\vect{Q}_1^{\star} = \vect{Q}_0 + h \vect{P}_{1/2} \\
\vect{P}_1^{\star} = \vect{P}_{1/2} -  \frac{h}{2} \nabla U(\vect{Q}_1^{\star}) 
\end{cases}
\end{equation} The `proposal move' $(\vect{Q}_1^{\star}, \vect{P}_1^{\star})$  is then accepted or rejected:
\begin{equation} \tag{Step 2}
\begin{pmatrix} \tilde{\vect{Q}}_1 \\ \tilde{\vect{P}}_1 \end{pmatrix} =  
x \begin{pmatrix} \vect{Q}_1^{\star} \\ \vect{P}_1^{\star} \end{pmatrix}  
+ (1-x) \begin{pmatrix} \vect{Q}_0 \\ - \vect{P}_0 \end{pmatrix} \;,
\end{equation}
where $x$ is a Bernoulli random variable with parameter $\alpha$, i.e., it takes value 1 with probability $\alpha$ and value 0 with probability $1-\alpha$.  The acceptance probability is defined as:
\begin{equation} \label{alpha} 
  \alpha =  1 \wedge \exp\left( - \frac{1}{kT} ( H(\vect{Q}_1^{\star}, \vect{P}_1^{\star})  - H(\vect{Q}_0, \vect{P}_0) ) \right) \;.
\end{equation}
The actual update of the system is taken to be: \begin{equation} \tag{Step 3}
\begin{pmatrix} \vect{Q}_1 \\
 \vect{P}_1 \end{pmatrix} = \begin{pmatrix} \tilde{\vect{Q}}_1 \\
  \exp(-\gamma h) \tilde{\vect{P}}_1 +  \sqrt{1 - \exp(-2 \gamma h)} \vect{m}^{1/2} \vect{\xi} \end{pmatrix} \;.
\end{equation}
Here $\vect{\xi} \in\mathbb{R}^n$ denotes a Gaussian random vector with mean zero and covariance $\E( \vect{\xi}_i \vect{\xi}_j ) = kT \vect{\delta}_{ij}$.
\end{algo}

Notice that the momentum gets flipped if a move is rejected in (Step 2).   This momentum flip is necessary in order to guarantee that the algorithm samples the correct stationary distribution \cite{AkBoRe2009}, but results in a $O(1)$ error in dynamics.   To compute dynamics not only must a long trajectory be stably produced with the right stationary distribution, but the approximation must also accurately represent the system's dynamics over the time interval of interest.   Unlike sampling algorithms, high acceptance rates are  needed to ensure that the time lag between successive rejections is frequently long enough to capture the desired dynamics.  Since the acceptance rate in \eqref{alpha} is related to how well the Verlet step preserves energy after a single step, this rejection rate is $O(h^{3})$.  Thus, in practice we find that the time-step required to obtain a $99.9 \%$ acceptance rate is often automatically satisfied with a time-step that sufficiently resolves the desired dynamics.  Each step of this algorithm requires: evaluating the atomic force field once in the third equation of (Step 1), generating a Bernoulli random variable with parameter $\alpha$ in (Step 2), and generating an $n$-dimensional Gaussian vector in (Step 3).  Since (Step 3) is the exact solution of \eqref{sdesplit1}, we stress that (Step 2) in Algorithm \ref{algo:integrator} is all that is really needed in most MD integration schemes to ensure that the integrator preserves the correct stationary density \eqref{langevindensity}.

Listing~\ref{list:integrator} translates Algorithm~\ref{algo:integrator} into the MATLAB language.  Intrinsically defined MATLAB functions appear in boldface.  The algorithm uses MATLAB's built in random number generators to carry out (Step 2) and (Step 3).  In particular, the Bernoulli random variable $x$ in (Step 2) is generated in {\em Line 15}, and the Gaussian vector in (Step 3) is generated on {\em Line 24}.   In addition to updating the positions and momenta of the system, the program also stores the previous value of the potential energy and force, so that the force and potential energy is evaluated just once in {\em Line 12} per simulation step.  This evaluation calls a MEX function which inputs the current position of the molecular system and outputs the force field and potential energy at that position.   We use a MEX function  because the atomistic force-field evaluation cannot be easily vectorized, and is by far, the most computationally demanding step in MD.  The {\tt PreProcessing} script file called in {\em Line 2} defines the physical and numerical parameters, sets the initial condition, and allocates space for storing simulation data.  Sample averages are updated as new points on the trajectory are produced in the {\tt UpdateSampleAverages} script file invoked in {\em Line 30}.  Finally, the outputs produced by the algorithm are handled by the {\tt PostProcessing} script file in {\em Line 34}.


\begin{lstlisting}[float, caption={Analysis-Based MD Integrator: {\tt MDintegrator.m}}, label=list:integrator]

PreProcessing;

for i = 1:Ns

    %--- Step 1 --- Velocity Verlet Proposal
    
    Ppt5=P0+0.5*h*F0;
    Q1star=Q0+h*Ppt5;
    [F1star,U1star]=ForceFieldmex(Q1star,Nm,rcut2,ell);
    P1star=Ppt5+0.5*h*F1star;
    
    %--- Step 2 --- Accept or Refuse Step 

    x=(rand<exp(-(0.5*P1star'*P1star-0.5*P0'*P0+U1star-U0)/kT));
    
    tildeP1=x*P1star-(1-x)*P0;
    tildeQ1=x*Q1star+(1-x)*Q0;
    F1=x*F1star+(1-x)*F0; U1=x*U1star+(1-x)*U0;

    %--- Step 3 --- Heat Bath Step
   
    Q1=tildeQ1;
    P1=f1*tildeP1+f2*randn(3*Nm,1);
    
    %--- iterate 
    
    Q0=Q1; P0=P1; F0=F1; U0=U1; 
    
    UpdateSampleAverages;
    
end

PostProcessing;
\end{lstlisting}


Let us consider a concrete example: a Lennard-Jones fluid that consists of $N$ identical atoms \cite{AlTi1987, FrSm2002, rapaport2004art}.  The configuration space of this system is a fixed cubic box with periodic boundary conditions.   The distance between the ith and jth particle is defined according to the {\em minimum image convention}, which states that the distance between $\vect{q}_i$ and $\vect{q}_j$ in a cubic box of length $\ell$ is:
\begin{equation} 
d_{MD}(\vect{q}_i, \vect{q}_j) \eqdef | ( \vect{q}_i - \vect{q}_j)  - \ell \lfloor ( \vect{q}_i - \vect{q}_j)/\ell \rceil  | \;.
\end{equation}
where $\lfloor \cdot \rceil$ is the nearest integer function.   In terms of this distance, the total potential energy is a sum over all pairs:
\begin{equation} \label{LJ}
U(\vect{q}) =  \sum_{i=1}^{n-1} \sum_{j=i+1}^{n} U_{LJ}(d_{MD}(\vect{q}_i, \vect{q}_j) ) \;,
\end{equation}
where $U_{LJ}(r)$ is the following truncated Lennard-Jones potential function:
\begin{equation} \label{LJpair}
U_{LJ}(r) = \begin{cases} f(r) - f(r_c) & r < r_c \;, \\
0 & \text{otherwise}  \;.
\end{cases}
\end{equation}
Here, $f(r) = 4 ( 1/r^{12} - 1/r^{6} )$ and $r_c$ is the cutoff radius which is bounded above by the size of the simulation box; and we have used dimensionless units to describe this system, where energy is rescaled by the depth of the Lennard-Jones potential energy and length by the point where the potential energy is zero.   The error introduced by the truncation in~\eqref{LJpair} is proportional to the density of the molecular system and can be made arbitrarily small by selecting the cutoff distance sufficiently large.  
Unless a neighbor and/or cell list is employed, a direct evaluation of the potential force $\nabla U(\vect{q})$ scales like $O(N^2)$, and typically dominates the total computational cost \cite{yao2004improved}.     Since the molecular system we consider will have just a few hundred atoms, we found that there is no advantage to using a fast force--field evaluation, and thus {\tt ForceFieldmex} evaluates the force and energy using a sum over all particle pairs.

\noindent
\begin{table}[hb!]
\centering
\begin{tabular}{|c|c|c|}
\hline
\multicolumn{1}{|c|}{\bf Parameter} & \multicolumn{1}{|c|}{\bf Description} &  \multicolumn{1}{|c|}{\bf Value}   \\
\hline
\hline
\multicolumn{3}{|c|}{ {\em Physical Parameters} } \\
\hline
\hline
$\rho$ & density & $\{0.6, 0.7, 0.8, 0.9, 1.0, 1.1 \}$ \\
  \hline
$kT$ & temperature & $0.5$  \\
  \hline
  $\gamma$ & heat bath parameter & $0.01$  \\
  \hline
$N_m$ & \# of molecules & $512$  \\
\hline
$T$ & time-span for autocorrelation & $2$ \\
\hline
\hline
\multicolumn{3}{|c|}{{\em  Numerical Parameters }} \\
\hline
\hline
$h$ & time-step & $0.005$ \\
\hline
$N_s$ & \# of simulation steps & $10^5$ \\
\hline
$r_{c}$ & LJ cutoff radius &  $2^{1/6}$ \\
\hline
\end{tabular}
\caption{ \small {\bf Simulation Parameters.}   
}
\label{simulationparameters}
\end{table}

\begin{figure}[ht!]
\includegraphics[width=\textwidth]{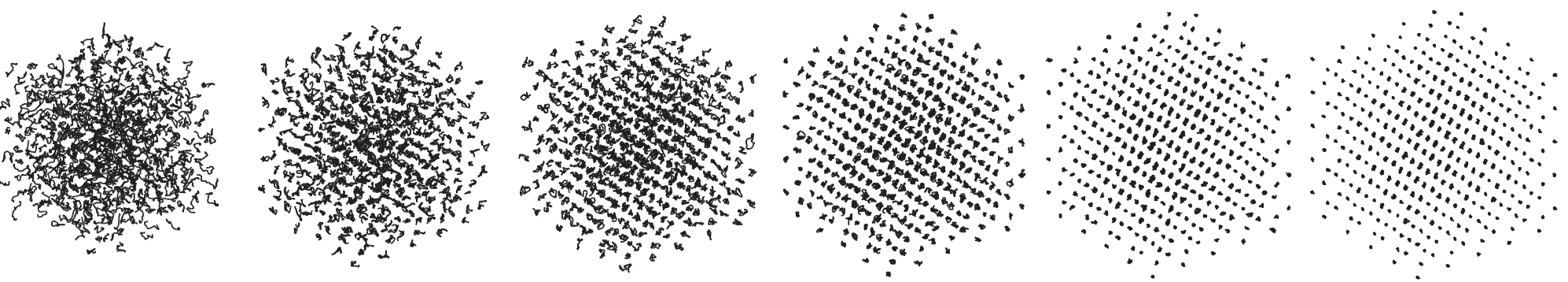} 
\hbox{
           \hspace{0.05in} (a) $\rho=0.6$
        \hspace{0.22in} (b) $\rho=0.7$
                \hspace{0.22in} (c) $\rho=0.8$
               \hspace{0.23in} (d) $\rho=0.9$ 
               \hspace{0.23in} (e) $\rho=1.0$
               \hspace{0.25in} (f) $\rho=1.1$
        } 
\caption{ \small {\bf Atomic Trajectories in Simulation Box.}   Plot of individual atomic trajectories from an initial condition where atoms are placed on the sites of an FCC lattice and at rest.  The trajectory is computed using the numerical and physical parameters indicated in Table~\ref{simulationparameters}, with the exception of the \# of steps which is set equal to $N_s = 1000$.   At the lower densities the particle trajectories are more diffusive and less localized.  
}
\label{fig:atomictrajectories}
\end{figure}

\begin{figure}[ht!]
\includegraphics[width=\textwidth]{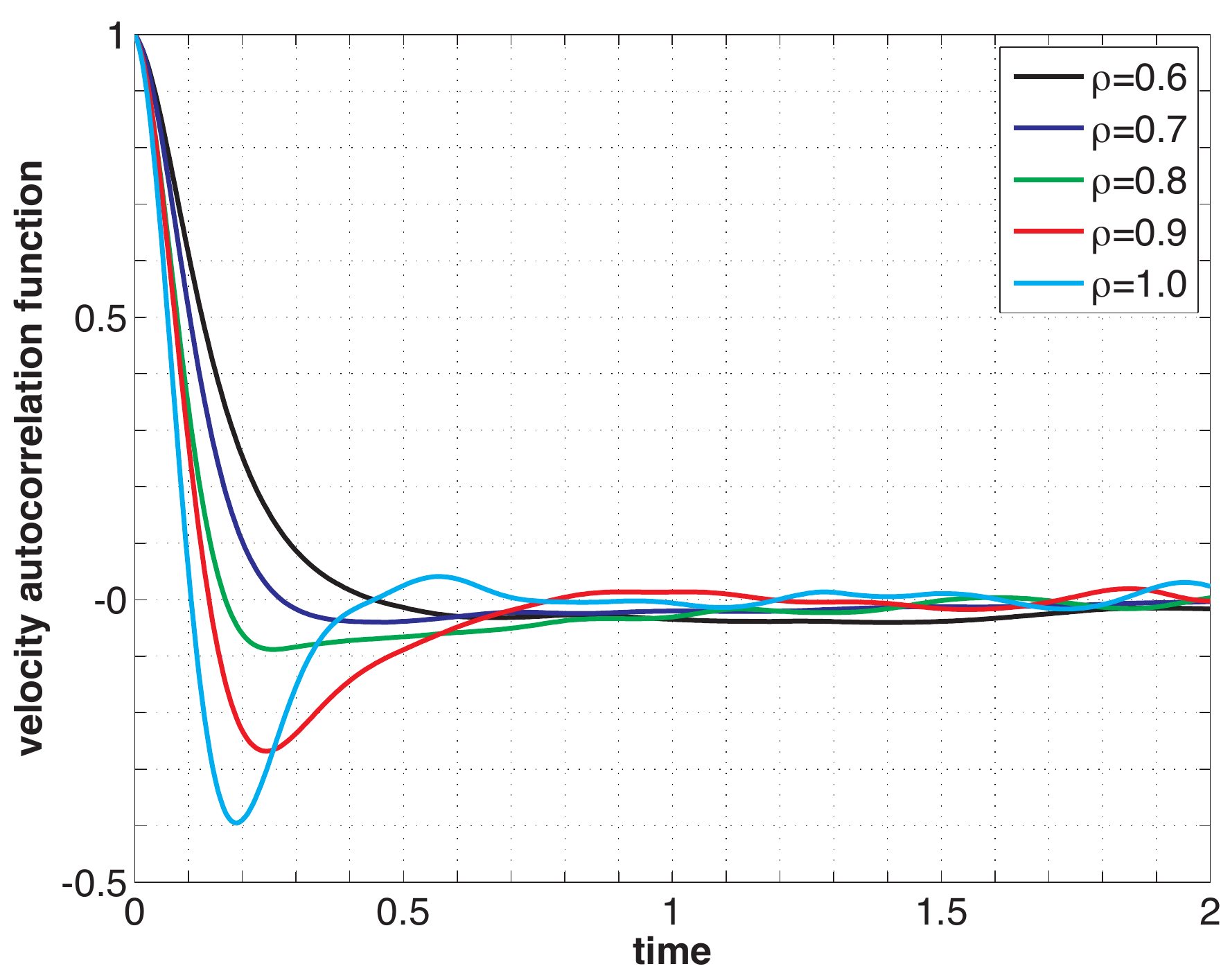} 
\caption{ \small {\bf Soft--sphere Velocity Autocorrelation Functions.}   A reproduction of Figure 5.2 of \cite{rapaport2004art} using Langevin dynamics with heat bath parameter $\gamma=0.01$.  The remaining parameters are set equal to those provided in Table~\ref{simulationparameters}.  The negative correlations at higher densities are consistent with what has been found in the literature \cite{rahman1964correlations,alder1967velocity}.  
 }
\label{fig:velocityautocorrelation}
\end{figure}

Listing~\ref{list:preprocessing} shows the {\tt PreProcessing} script which sets the parameters provided in Table~\ref{simulationparameters} and constructs the initial condition, where the $N$ atoms are assumed to be at rest and on the sites of an FCC lattice.   The command {\tt rng(123)} on {\em Line 3} sets the seed of the random number generator functions {\tt RAND} and {\tt RANDN}.   The acceptance rates at every step and the velocity autocorrelation are updated in the {\tt UpdateSampleAverages} script shown in Listing~\ref{list:updatesampleaverages}.   The mean acceptance rate which is outputted in the script {\tt PostProcessing} shown in Listing~\ref{list:postprocessing} must be high enough to ensure that the dynamics is accurately represented.   To compute the autocorrelation of an observable over a time interval of length $T$,  the value of that observable along the entire trajectory is not needed.  In fact, all that is really needed are the values of this observable along a piece of trajectory over the moving time-window $[t_i, t_i+T]$ where $t_i = i \times h$.  This storage space is allocated in {\tt PreProcessing} and is updated in {\tt UpdateSampleAverages}.  More precisely, since we define $N_a = \lceil T/ h \rceil + 1$ in the {\tt PreProcessing} script, the molecular velocities are stored in the {\tt pivot} array from $i-N_a$ to $i$, where $i$ is the index of the current position.   Notice that velocity autocorrelations are not computed until after the index $i$ exceeds $10^4$.  This {\em equilibration time} removes some of the statistical bias that may arise from using a non-random initial condition.  Short-time trajectories of this molecular system are plotted in Figure~\ref{fig:atomictrajectories}.  This figure shows that the particle motions are more localized at higher densities.  Using the parameters provided in Table~\ref{simulationparameters}, we compute velocity autocorrelations for a range of density values in Figure~\ref{fig:velocityautocorrelation}.   Since the heat bath parameter is set to a small value, these figures are in agreement with those obtained by simulating the molecular system with no heat bath in Figure 5.2 of \cite{rapaport2004art}.


\begin{lstlisting}[float, caption={Analysis-Based MD Integrator: {\tt PreProcessing.m}}, label=list:preprocessing]
%--- seed random # generator

rng(123);

%--- physical parameters

rho=0.6;                    % density
kT=0.5;                     % temperature factor
gama=0.1;                   % heat bath parameter
Nm=500;                     % # of molecules
T=2.0;                      % time span for velocity correlation
ell=(Nm/rho)^(1/3);         % length of cubic box

%--- simulation parameters

h=0.005;                    % time-step size
Ns=1e3;                     % # of steps
rcut = 2.0^(1/6);           % cutoff radius
rcut2 = rcut*rcut;
 
f1=exp(-gama*h); f2=sqrt((1.0-exp(-2.0*gama*h))*kT);

%--- initial condition 

A=fcclattice(Nm,ell);
Q0=reshape(A, [3*Nm 1]);    % atoms on an fcc lattice
P0=zeros(3*Nm,1);           % atoms at rest

%--- initialize statistics

NA=ceil(T/h)+1;             % preallocate space for 
acf=zeros(NA,1);            % online correlation computation
varacf=zeros(NA,1);
pivot=zeros(NA,3*Nm);
nacf=zeros(NA,1);

AP=zeros(Ns,1);             % vector of acceptance probabilities

[F0,U0]=ForceFieldmex(Q0,Nm,rcut2,ell);  % initial force & energy
\end{lstlisting}

\begin{lstlisting}[float, caption={Analysis-Based MD Integrator: {\tt UpdateSampleAverages.m}}, label=list:updatesampleaverages]
%--- store acceptance probability

AP(i)=x;

%--- update correlation function 

if (i>1e4)
        
    pp=mod(i-1,NA)+1;
    pivot(pp,:)=P0;
    
    for j=1:min(i,NA)
        nacf(j)=nacf(j)+1;
        mui=acf(j);
        vari=varacf(j);
        n_samples=nacf(j);
        xip1=pivot(mod(pp-j,NA)+1,:)*pivot(pp,:)'/(3.0*Nm);
        acf(j)=mui+(xip1-mui)/n_samples;
        varacf(j)=((n_samples-1)*vari+...
            (xip1-mui)*(xip1-acf(j)))/n_samples;
    end
    
end
\end{lstlisting}

\begin{lstlisting}[float, caption={Analysis-Based MD Integrator: {\tt PostProcessing.m}}, label=list:postprocessing]
%--- output results

disp(['h=' num2str(h)  ',<AP>=' num2str(mean(AP))]);

figure(2); clf; hold on; tt=0:h:T;
errorbar(tt,acf,1.96*sqrt(varacf)./sqrt(nacf)); 

save('VelocityAutocorrelation.mat', 'tt', 'acf', 'varacf');
\end{lstlisting}

\section{Potential Pitfalls in High Dimension \& Tricks for Mitigation} \label{scaling}

For high dimensional systems (think $N>10^3$ atoms), calculating the force-field at every step is the main computational cost of MD simulation.  These force fields involve: bonded interactions, and non-bonded Lennard-Jones \& electrostatic interactions.  The calculation of bonded interactions is straightforward to vectorize and scales like $O(N)$.  In addition, Lennard-Jones forces rapidly decay with interatomic distance.  To a good approximation, every atom interacts only with neighbors within a sufficiently large ball.  By using data structures like neighbor lists and cell linked-lists, these interactions can be calculated in $O(N)$ steps, and therefore, the Lennard-Jones interactions can be calculated in $O(N)$ steps \cite{yao2004improved}.  On the other hand, the electrostatic energy between particles decays like $1/r$ where $r$ denotes an interbead distance which leads to long-range interactions between atoms.  Unlike Lennard--Jones interaction, this interaction cannot be cutoff without introducing large errors.  In this case, one can use sophisticated analysis-based techniques like the fast multipole method to rigorously handle such interactions in $\mathcal{O}(N)$ steps \cite{greengard1987fast, weinan2011principles}.

However, the effect of these `mathematical tricks' for fast calculation of the force field can become muted if the time-step requirement for stability or accuracy becomes more severe in high dimension.   This can happen in the Metropolis integrator, if the acceptance probability in (Step 2) of Algorithm~\ref{algo:integrator} deteriorates in high dimension.   The scaling of Metropolis algorithms has been quantified for the random walk Metropolis, hybrid Monte-Carlo, and MALA algorithms \cite{GeGiRo1997,  RoRo1998, BeRoSt2009, BePiRoSaSt2010, MaPiSt2012}.  Since the acceptance probability is a function of an extensive quantity, the acceptance rate can artificially deteriorate with increasing system size unless the time-step is reduced.     Because high acceptance rates are required to maintain dynamic accuracy, the dependence of the time-step on system size limits the application of Metropolized schemes to large-scale systems.    Fortunately, this scalability issue can often be resolved by using local rather than global proposal moves because the change in energy induced by a local move is typically an intensive quantity.    For molecular dynamics calculations, this approach was pursued in \cite{BoVa2012}.  Using dynamically consistent local moves (a so-called J-splitting \cite{FeWa1994}), it was shown that in certain situations a scalable Metropolis integrator can be designed; however, the extent to which this strategy remedies the issue of high rejection rate in high dimension is not clear at this point, and should be tested in applications.


\section{Conclusion} \label{conclusion}

This paper provided a step-by-step algorithmic introduction to new {\em analysis-based} MD integrators for MD simulation.  These algorithms are long-time stable and finite-time accurate for MD simulation.   A MATLAB implementation of the algorithm was provided, and used to compute the velocity autocorrelation of a sea of Lennard--Jones particles at various densities between the solid and liquid phases.   The paper did not review the theory of Metropolis integrators which is discussed elsewhere \cite{BoVa2010, BoVa2012}.  We conclude by mentioning two directions for future development.
\begin{itemize}
\item
The compatibility of the Metropolis integrator with a fast multipole method for handling the electrostatic effects has not been investigated.  This loose end is indicated by the question mark appearing towards the right end of the timeline in Figure~\ref{fig:timeline}, which points out that an $O(N)$ Metropolis integrator is currently unavailable. 
\item
Another issue is the development of second-order Metropolis integrators for \eqref{langevin} which might permit larger time-steps.  The `quasi-symplectic' second--order Langevin integrators introduced in \cite{VaCi2006} would be a natural starting point for designing proposal moves that would lead to second-order Metropolis integrators.
\end{itemize}


\acknowledgements{Acknowledgements}

The research that led to this paper was made possible by NSF grant DMS-1212058. 


\conflictofinterests{Conflict of Interest}

The authors declare no conflict of interest. 

\bibliographystyle{mdpi}
\bibliography{nawaf}

\end{document}